\documentclass[aps,reprint,groupaddress,superscriptaddress,amssymb,prx]{revtex4-2}
\usepackage[utf8]{inputenc}
\usepackage{color}
\usepackage{xcolor}
\usepackage{graphicx}
\usepackage{stmaryrd}

\usepackage{mdframed}
\usepackage{physics}
\usepackage{amsmath}
\usepackage{amsfonts}
\usepackage{bm}
\usepackage[symbol]{footmisc}

\begin{document}
\preprint{APS/123-QED}

\title{
Information propagation in multilayer systems\\with higher-order interactions across timescales}

\author{Giorgio Nicoletti}
\affiliation{Laboratory of Interdisciplinary Physics, Department of Physics and Astronomy ``Galileo Galilei'', University of Padova, Padova, Italy}
\affiliation{Department of Mathematics ``Tullio Levi-Civita'', University of Padova, Padova, Italy}
\affiliation{ECHO Laboratory, École Polytechnique Fédérale de Lausanne, Lausanne, Switzerland}
\author{Daniel Maria Busiello}
\affiliation{Max Planck Institute for the Physics of Complex Systems, Dresden, Germany}

\begin{abstract}
\noindent Complex systems are characterized by multiple spatial and temporal scales. A natural framework to capture their multiscale nature is that of multilayer networks, where different layers represent distinct physical processes that often regulate each other indirectly. We model these regulatory mechanisms through triadic higher-order interactions between nodes and edges. In this work, we focus on how the different timescales associated with each layer impact their effective couplings in terms of their mutual information. We unravel the general principles governing how such information propagates across the multiscale structure, and apply them to study archetypal examples of biological signaling networks and effective environmental dependencies in stochastic processes. Our framework generalizes to any dynamics on multilayer networks, paving the way for a deeper understanding of how the multiscale nature of real-world systems shapes their information content and complexity. 
\end{abstract}

\maketitle
    
\vspace{0.5cm}

\noindent Real-world systems are often characterized by interconnected dynamical processes occurring at multiple scales, both spatial and temporal ones. This multiscale structure is a fundamental ingredient in shaping the properties of complex systems \cite{dedomenico2023more}. Further, such intrinsic temporal separation and the interplay between slow and fast processes have been known to be a crucial feature of biological, chemical, and ecological systems \cite{schaffer2021mapping, hastings2010timescales, radulescu2008robust, poisot2015beyond, timme2014multiplex, cavanagh2020diversity, honey2017switching, melykuti2014equilibrium, nicoletti2023emergence}.

A natural framework to describe these interconnected multiscale systems is that of multilayer networks \cite{bianconi2018multilayer, dedomenico2013mathematical, boccaletti2014structure, artime2022multilayer, aleta2019multilayer}. Nodes in a layer may represent either physical or biological units, habitats on complex landscapes, or states in a chemical system, with each layer embedding different processes - from random walk, agent-based, and master equation dynamics, to spreading and stochastic processes \cite{salehi2015spreading, dedomenico2016physics, dearruda2018fundamentals, ghavasieh2020statistical, flatt2023abc, nicoletti2023emergent}. Here, we consider the different layers as physically separated processes that take place at distinct timescales and regulate each other indirectly. In this scenario, as we will detail later on, interactions across timescales are not pairwise, but rather higher-order in nature. Higher-order interactions \cite{battiston2020networks, battiston2021physics} and their effects on a variety of phenomena have been extensively studied in recent years, from synchronization to epidemic spreading and the stability of ecological systems \cite{bairey2016high, grilli2017higher, letten2019mechanistic, carletti2020random, millan2020explosive, stonge2021universal, majhi2022dynamics, gibbs2022coexistence}. In particular, the higher-order interactions considered herein represent how a node evolving with a given timescale regulates the pairwise link between two nodes evolving with another timescale. These kinds of interactions are known as triadic interactions \cite{sun2023dynamic}. This setting is especially relevant in neuroscience, ecology, and climate science \cite{bairey2016high, grilli2017higher, letten2019mechanistic, cho2016optogenetic, boers2019complex, faskowitz2022edges}, but it has also been used to characterize the control of stochastic reaction networks \cite{hilfinger2016constraints,yan2019kinetic}, a general framework with applications in signaling and regulatory biochemical processes \cite{ma2009defining, rahi2017oscillatory}.

Here, we derive a general decomposition for the probability distribution of any dynamical process taking place in such multilayer networks with triadic interactions between layers, i.e., timescales. By inspecting this decomposition, we quantify how the higher-order structure couples the layers by computing their mutual information exactly, and we unravel the principles governing how this information propagates across timescales. To prove the versatility of our framework, we employ it to study biological signaling networks and to show how disjointed components of a fast layer become effectively coupled due to shared slow interactions, an archetypal model for the interplay between environmental and internal dependencies in complex systems.

\section*{Information across timescales}
\begin{figure}
    \centering
    \includegraphics[width = \columnwidth]{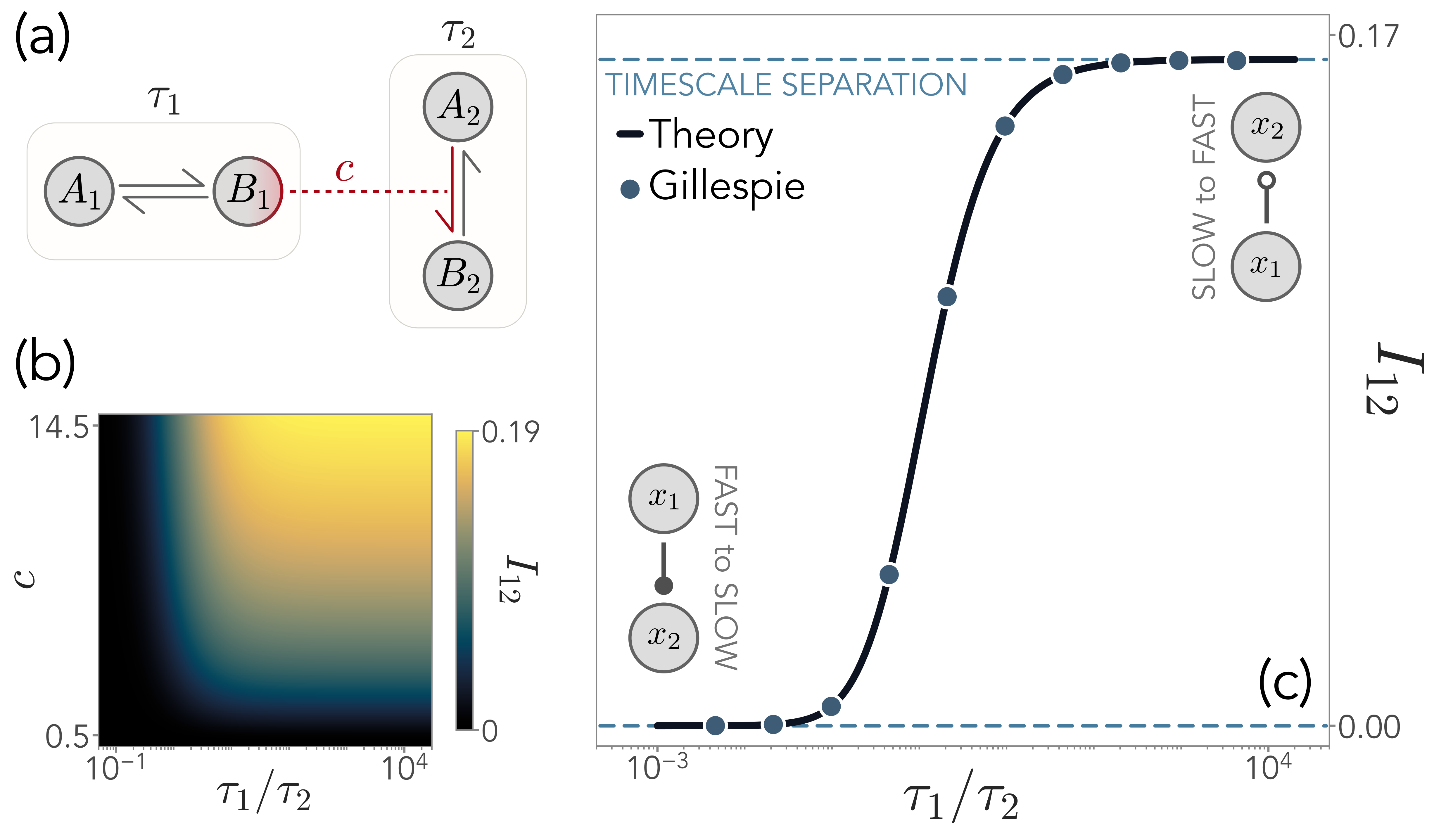}
    \caption{Information between two layers with states $x_1$ and $x_2$ is influenced by the their timescale ratio $\tau_1/\tau_2$. (a) Sketch of the system with the node $B_1$ influencing the link $A_2 \to B_2$. (b) Mutual information $I_{12}$ between $x_1$ and $x_2$ as a function of $\tau_1/\tau_2$ and the triadic interaction strength $c$. (c) At fixed $c = 10$, $I_{12}$ vanishes when $x_1$ is faster than $x_2$. When $\tau_1/\tau_2 \to \infty$, the regulating layer is slower, and $I_{12}$ converges to a non-zero value. The master equation of the system can be solved exactly (solid black line), simulated via a Gillespie algorithm (blue dots), and solved in a timescale separation regime (dashed blue line).}
    \label{fig:mutual_tau}
\end{figure}

\noindent As a proof of concept, we first consider the simple case of a two-layer system. Each layer has two nodes, $A_\mu$ and $B_\mu$, with $\mu = 1, 2$, and two directed links connecting them with weights $w_\mu^{A \to B}/\tau_\mu$ and $w_\mu^{B \to A}/\tau_\mu$ (see Figure \ref{fig:mutual_tau}a). Here, $\tau_\mu$ is the characteristic timescale of the layer dynamics. We name $x_\mu$ the state of the layer $\mu$, which can thus take two possible values, $A_\mu$ or $B_\mu$. We can think of $x_\mu$ as describing a random walk in each layer, or equivalently a molecule switching between two possible configurations, a two-state chemical network, or any other master equation dynamics \cite{gardiner}.

The two layers represent two stochastic systems physically separated from each other. In other words, a random walker moving in the first layer cannot jump on the second one, or a molecule cannot switch between states belonging to different layers. As a consequence, the layers are causally connected via triadic interactions, so that the state of a node in the first layer influences an edge in the second one (dashed line in Figure \ref{fig:mutual_tau}a). For simplicity, we write the overall transition rate $\omega_2^{A \to B}$ as
\begin{equation}
\label{eqn:triadic_int_twolayers}
    \omega_2^{A \to B} = \frac{w_2^{A \to B} + c\, \delta(x_1, B_1)}{\tau_2}
\end{equation}
where $\delta(\cdot)$ is the Kronecker delta and $c$ represents the regulatory interaction that stimulates the transitions to $B_2$ when the first random walker is in the state $B_1$.

We can solve the master equation for the joint probability of the two layers exactly, and study the mutual information $I_{12}$ between them (see Methods). In Figure \ref{fig:mutual_tau}b-c, we show the behavior of $I_{12}$ as a function of the triadic interaction strength $c$ and the relative timescale separation between the two layers, $\tau_1/\tau_2$. As expected, $I_{12}$ increases with $c$, since when $c=0$, the two layers are decoupled. Additionally, we find that no information is shared between the two layers when $\tau_1 \ll \tau_2$, i.e., when the regulating layer - the first one - evolves faster than the regulated layer - the second one. However, information is maximal when $\vec{x}_1$ undergoes a much slower dynamics than $\vec{x}_2$, i.e., $\tau_1 \gg \tau_2$, with a steep transition between these two regimes.

\begin{figure}[b]
    \centering
    \includegraphics[width = \columnwidth]{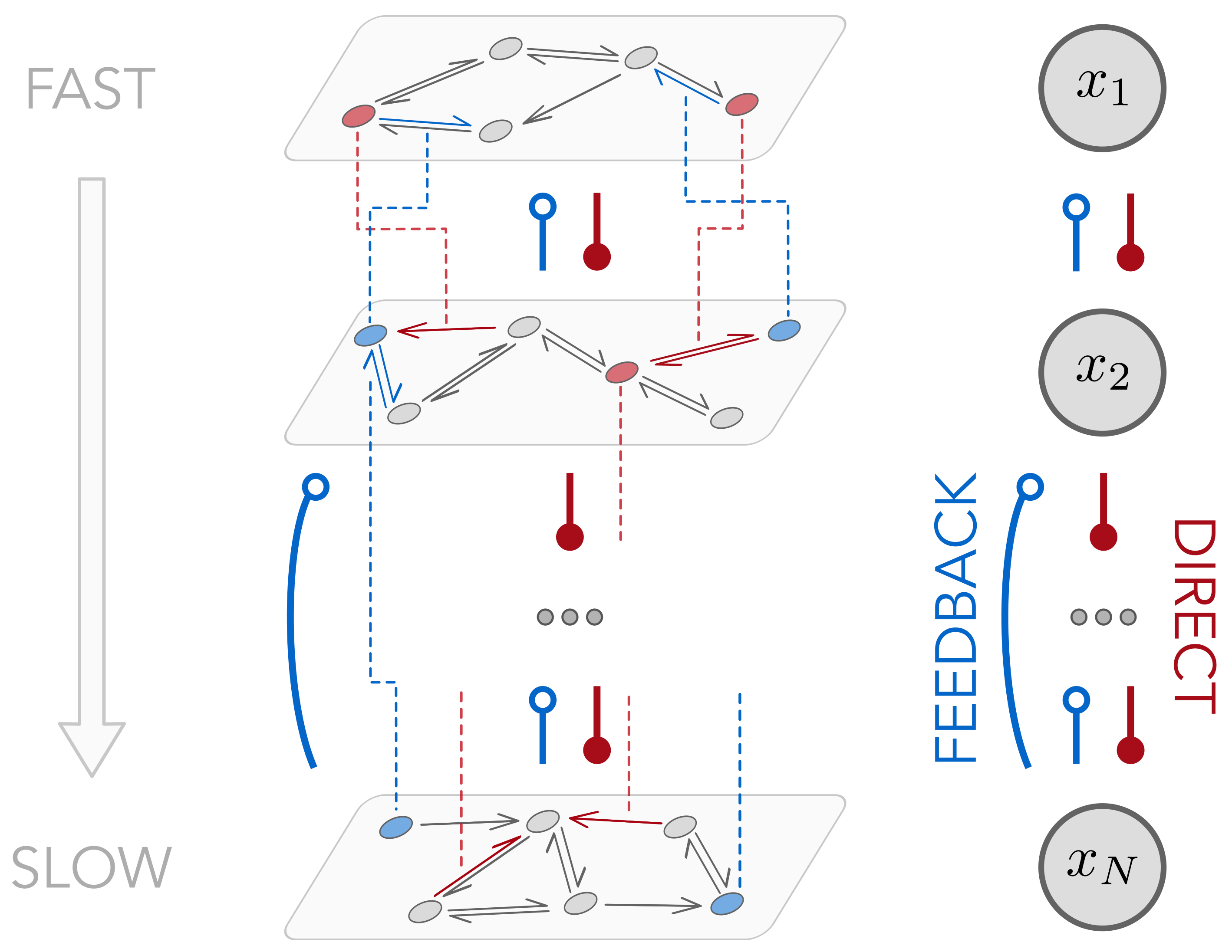}
    \caption{Sketch of a multilayer system with intra-layer pairwise interactions and interactions across layers. We name inter-layer interactions going from a layer to a slower one as ``direct'' interactions, as they follow the ordering induced by the timescales. On the contrary, we refer to those that go towards faster layers as ``feedback'' interactions.}
    \label{fig:sketch}
\end{figure}

This simple example suggests that the relative characteristic timescales among physically separated systems shape the information content shared by different layers. Our goal is to unveil this connection on more general grounds and, therefore, the main principles governing information propagation in multilayer systems.

\begin{figure*}[t]
    \centering
    \includegraphics[width = \textwidth]{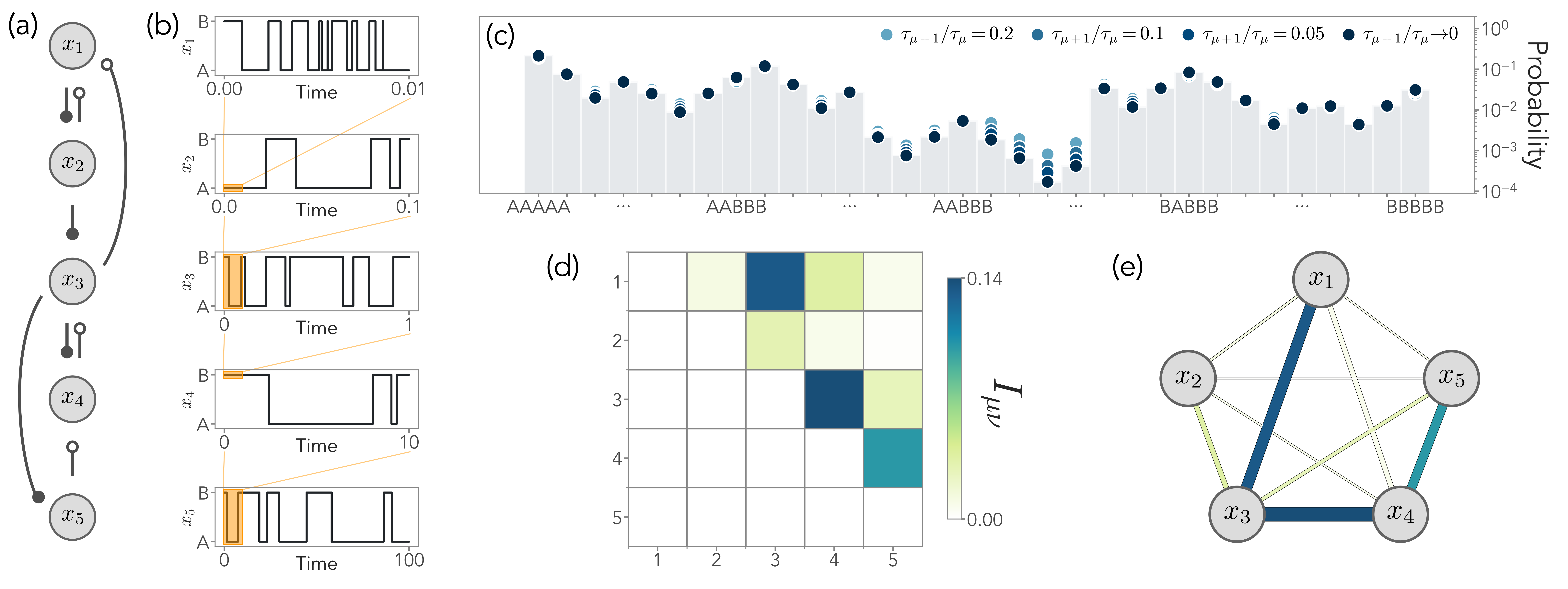}
    \caption{Accuracy of the timescale separation solution and mutual information in a five-layer system with triadic interactions. (a) Scheme of the interactions, defined in Eq.~\eqref{eqn:rate_matrix}. For simplicity, we set feedback interactions as $F_{\mu\nu}^{k, i \to j} = F_{\mu\nu}^\mathrm{eq}\delta_{jk} + F_{\mu\nu}^\mathrm{cr}(1-\delta_{jk})$, and similarly for direct interactions $D_{\mu\nu}^{k, i \to j}$, with $F_{\mu\nu}^\mathrm{eq} = 10$, $D_{\mu\nu}^\mathrm{eq} = 0.1 = F_{\mu\nu}^\mathrm{cr}$, and $D_{\mu\nu}^\mathrm{cr} = 5$ (see Methods). (b) Simulated trajectories in each layer with $\tau_{\mu + 1}/\tau_\mu = 0.1$, so that each layer is an order of magnitude faster than the previous one, using a Gillespie algorithm. (c) Comparison of the joint probability estimated from simulated trajectories at different separations between the timescales of each layer. Grey bars represent the analytical solution in Eq.~\eqref{eqn:joint_probability}, which is well approximated already for $\tau_{\mu + 1}/\tau_\mu = 0.1$. (d) MIMMO associated with this system when $\tau_{\mu + 1}/\tau_\mu \to 0$. (e) Mutual information between each pair of layers is shaped by the pathways among them defined by direct and feedback interactions.}
    \label{fig:gillespie_infonet}
\end{figure*}

\section*{General multilayer dynamics}
\noindent Consider a system of $N$ layers. The $\mu$-th layer is a network with $M_\mu$ nodes, defined by a $M_\mu \times M_\mu$ adjacency matrix $\hat{W}_\mu$ that encodes the pairwise relations between nodes, e.g., transition rates or interactions. For easiness of notation, here we focus on the case in which each layer supports a random walk or master equation dynamics, so that we can identify the state of a layer with the occupied node $x_\mu$, as in the previous example. However, our framework is valid for any stochastic dynamics on networks, from Ornstein-Uhlenbeck processes to models of neural dynamics, where the state of a layer is a vector of node states \cite{gardiner, mariani2022disentangling}.

Causal interactions across layers are triadic, going from nodes of one layer to edges of another. As outlined above, this setting naturally stems from the requirement that layers represent physically separated systems. Such higher-order interactions are defined by a tensor $C_{\mu \nu}^{k, i \to j}$, which denotes how the node $k$ in layer $\mu$ influences the $i \to j$ interaction in layer $\nu$. The tensor $\hat{C}_{\mu\nu}$ then contains the set of all possible interactions from $\mu$ to $\nu$. Thus, the overall adjacency matrix $\hat\Omega_\mu$ of a layer will depend on both the intra-layer pairwise interactions between the nodes, as well as on inter-layer interactions coming from the nodes of other layers. To fix the ideas, when triadic interactions are additive with respect to pairwise couplings, we write its elements as
\begin{equation}
\label{eqn:rate_matrix}
    \hat\Omega_\mu\left(\{x\}_{\rightsquigarrow \mu}\right)^{i \to j} = (\hat{W}_\mu)^{i \to j} + \sum_{\nu \ne \mu} \sum_{k = 1}^{M_\nu} C_{\nu\mu}^{k, i \to j} \phi_\mu^k(x_\nu) \;.
\end{equation}
In Eq.~\eqref{eqn:rate_matrix}, $\phi_\mu^k(x_\nu)$ is a generic non-linear function of the state of a layer, and $\{x\}_{\rightsquigarrow \mu}$ is a shorthand notation to denote the state of all layers connected to $\mu$ by higher-order interactions, i.e., all $\nu$ for which $\hat{C}_{\nu\mu}$ is non-zero. Let us stress that the additive structure in Eq.~\eqref{eqn:rate_matrix} is only exemplary, since our results are independent of the specific form of $\hat{\Omega}_\mu$, where triadic interactions may enter in any functional shape.

\begin{figure*}[t]
    \centering
    \includegraphics[width = 0.9 \textwidth]{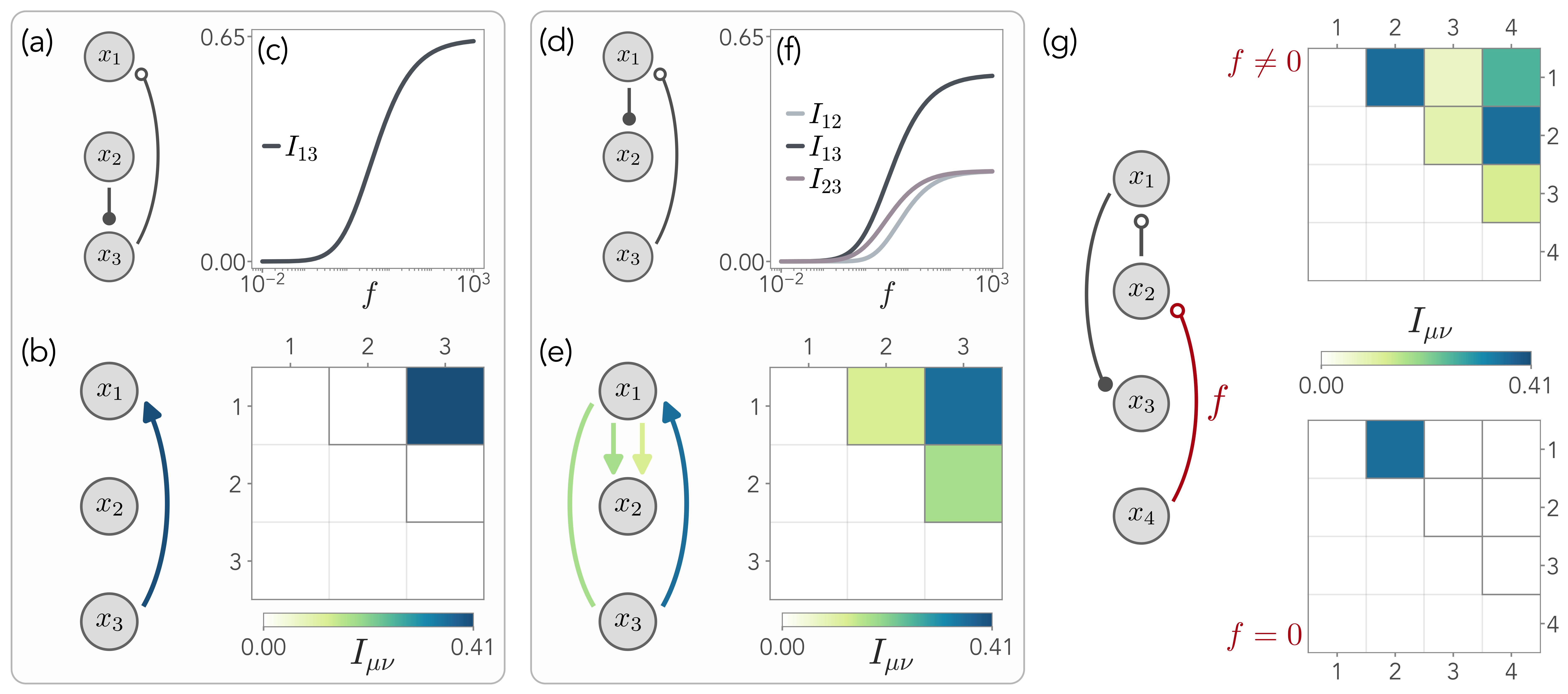}
    \caption{The principles of information propagation in multilayer networks. (a) A three-layer system with a higher-order feedback interaction from layer $3$ to $1$ and a direct link from $2$ to $3$. Each layer has two nodes, as in the previous examples. (b) The slow layer acts as an information source for the fast one so that $I_{13} \ne 0$ in the MIMMO, while the direct link does not generate information. (c) A stronger feedback interaction $f$ increases the mutual information between the two layers. (d) The same system as the first panel, with the direct link from $1$ to $2$. (e) All the elements of the MIMMO are non-zero, since it exists a minimal propagation path $3 \to 1 \to 2$. The direct link propagates the information created in $1$. (f) All the entries of the MIMMO depend on $f$, and they vanish as $f$ goes to zero, i.e., no information is created. (g) A four-layer system with feedback interactions from $4$ to $2$ and $2$ to $1$, and a direct link from $1$ to $3$. While for $f\neq 0$ all the entries of the MIMMO are non-zero, when $f = 0$, the information created in $1$ by $2$ cannot be propagated to $3$, since the dynamics of $x_3$ is faster than $x_2$. As a consequence, only $I_{12} \neq 0$.}
    \label{fig:information_propagation}
\end{figure*}

We now order the layers by their timescales - that is, we take $\tau_1 < \tau_2 < \dots < \tau_N$ so that the first layer is the fastest and the last the slowest. We name triadic interactions going from a fast to a slow layer as ``direct'' interactions. Vice-versa, when a slow node influences a fast link, we call the corresponding interaction a ``feedback'' one. Hence, the tensor containing all causal interactions, $\hat{C}_{\mu\nu}$, can be written as the sum of two contributions,
\begin{equation}
\label{eqn:feedback_direct}
    \hat{C}_{\mu \nu} = \underbrace{\hat{D}_{\mu \nu}}_{\mu < \nu} + \underbrace{\hat{F}_{\mu \nu}}_{\mu > \nu}
\end{equation}
where $D_{\mu \nu}$ is a lower-triangular tensor in the indices $\mu$ and $\nu$ and contains direct interactions, whereas $F_{\mu \nu}^{k, i \to j}$ is upper-triangular and describes feedback interactions. Intuitively, these two kinds of interactions capture different physical mechanisms: direct couplings act as controls modulated by underlying fast processes, while feedback interactions might encapsulate regulatory schemes. As such, as we will see, they play a very different role in determining how information is created and propagated across layers. This general model is depicted in Figure \ref{fig:sketch}.

The joint probability distribution of the states of each layer at time $t$, $p_{1, \dots, N}(t)$ obeys the master equation
\begin{equation}
\label{eqn:master_equation}
    \partial_t p_{1, \dots, N}(t) = \sum_{\mu = 1}^N \frac{\hat{\Omega}_\mu(\{x\}_{\rightsquigarrow \mu})}{\tau_\mu} \, p_{1, \dots, N}(t) \;.
\end{equation}
Inspired by the results of the previous section, we are particularly interested in the timescale separation regime $\tau_{\mu+1} / \tau_\mu \to 0$ \cite{bo2017multiple}. The presence of both direct and feedback interactions allows us to explore both the extreme behaviors - maximum and zero information - highlighted in the previous section. In this regime, we can solve Eq.~\eqref{eqn:master_equation} analytically (see Methods) and show that $p_{1,\dots,N}$ takes the following factorized form:
\begin{equation}
\label{eqn:joint_probability}
    p_{1, \dots, N}(t) = p_N^\mathrm{eff}(t) \prod_{\nu = 1}^{N-1}p_{\mu | \rho(\mu)}^{\rm st, eff} \;.
\end{equation}
Here, $\rho(\mu)$ appearing in the conditional probabilities is the set of all layers interacting with $\mu$ via either a single feedback link or a minimal propagation path (mPP). Any propagation path (PP) to $\mu$ is a directional path coming from layers slower than $\mu$ and containing at least one direct link. An mPP from $\nu$ to $\mu$ is then a PP on the induced subgraph obtained by removing all layers slower than $\mu$ except $\nu$ (see Methods).
Finally, $p_{\mu | \rho(\mu)}^{\rm st, eff}$ is the stationary probability of the effective operator
\begin{equation}
\label{eqn:effective_operators}
    \hat{\Omega}^\mathrm{eff}_{\mu | \rho(\mu)} = \sum_{x_1} \dots \sum_{x_{\mu -1}} \hat{\Omega}_\mu\left(\{x\}_{\rightsquigarrow \mu}\right) \prod_{\nu = 1}^{\mu - 1} p_{\nu | \rho(\nu)}^{\rm st, eff}
\end{equation}
that averages the effect of all the layers faster than $\mu$ and explicitly depends on those slower than $\mu$ belonging to $\rho(\mu)$. Clearly, $p^{\rm st, eff} = p^{\rm st}$ if there is no direct link towards the layer $\mu$. Hence, the timescale separation defines a hierarchy of effective operators and their stationary probabilities, starting from the fastest layer. In terms of these operators, the multilayer joint probability can be written exactly as a product of suitable conditional probabilities, and the effective operator of the slowest layer determines its time evolution. From the stationary joint probability $p_{1, \dots, N}^\mathrm{st}$, we are able to compute all marginal probabilities and thus the mutual information between the layers:
\begin{equation}
\label{eqn:MIMMO}
    I_{\mu\nu} = \sum_{x_\mu, x_\nu} p_{\mu\nu} \,\log_2\frac{p_{\mu\nu}}{p_\mu \,p_\nu}
\end{equation}
defining a symmetric matrix that we name Mutual Information Matrix for Multiscale Observables (MIMMO).

Despite its intricate structure, the proposed framework provides us with a very instructive and intuitive way to deal with multilayer systems operating at different timescales. To show this beyond the general theory, we study the example of the five-layer system in Figure \ref{fig:gillespie_infonet}a. Each layer has two nodes $\mathcal{N} = \{A, B\}$. Triadic interactions between layers are additive, as in Eq~\eqref{eqn:rate_matrix}, with $\phi_\mu^k (x_\nu) = \delta(x_\nu, k)$ for all layers $\mu$, with $k \in \mathcal{N}$. In this way, they are switched on and off depending on the state of the regulating layer (see Methods for details). Figure \ref{fig:gillespie_infonet}b shows the typical trajectories obtained through an exact Gillespie algorithm \cite{gillespie1977exact}. From Eq.~\eqref{eqn:joint_probability}, the joint probability distribution reads: $p_{1,2,3,4,5}(t) = p^{\rm st}_{1|2,3} \, p^{\rm st, eff}_{2} \, p^{\rm st, eff}_{3|4} \, p^{\rm st, eff}_{4|5} \, p^{\rm eff}_5(t)$. We can compare this analytical solution with simulations at different values of $\tau_{\mu + 1}/\tau_\mu$. As we see in Figure \ref{fig:gillespie_infonet}c, the timescale separation assumption provides an excellent approximation for the system even when the timescales differ by just one order of magnitude, i.e., $\tau_{\mu + 1}/\tau_\mu  = 0.1$. Finally, we show the upper triangular section of the Mutual Information Matrix in Figure \ref{fig:gillespie_infonet}d. We remark that, even for $\tau_{\mu + 1}/\tau_\mu = 0.1$, the MIMMO obtained directly from trajectories is qualitatively identical to the analytical one. From the MIMMO, we can assign an undirected link between any two layers $\mu$ and $\nu$ with a weight equal to $I_{\mu\nu}$, resulting in a fully connected network that quantifies how much information each pair of layers share (Figure \ref{fig:gillespie_infonet}e).

Already at the level of this paradigmatic example, it is evident that direct and feedback interactions both play a role in determining how information spreads among layers due to their causal relationships, i.e., how much layers are coupled to one another. To unravel the differences between these two classes of interactions, we now exploit the conditional structure of the complete solution.

\begin{figure*}
    \centering
    \includegraphics[width = 0.9 \textwidth]{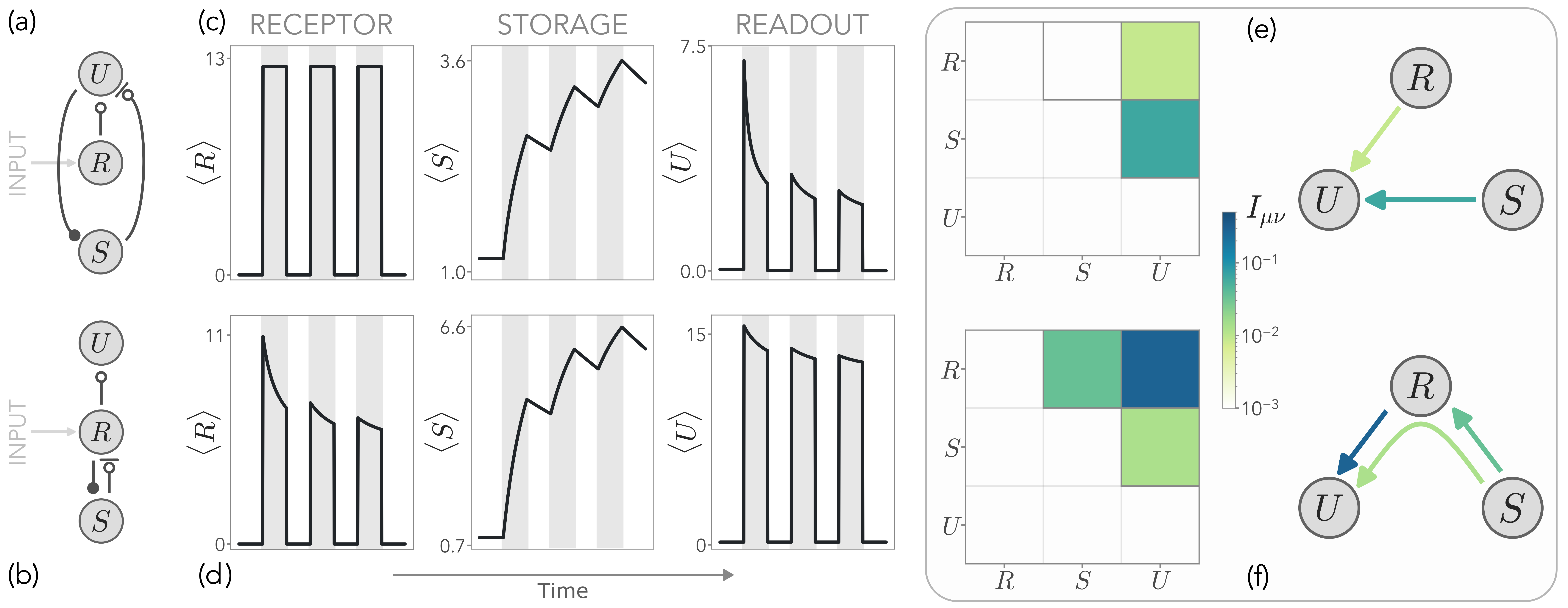}
    \caption{Information in birth-and-death processes modeling multi-scale signaling networks. (a-b) Two biochemical architectures known to implement biological adaptation, encompassing a receptor $R$, a storage $S$, and a readout population $U$. Inhibitory links, which increase the death rate of a population, are indicated with a bar at the end. (c-d) Both architectures display similar dynamical behavior in the presence of a repeated input to the receptor, with the readout average $\ev U$ decreasing and $\ev S$ increasing at each signal (gray shaded area). (e) The two architectures display important differences at the information level across layers. In the first, the receptor and the storage share no mutual information, and the regulation of $S$ acts independently of the signal. (f) In the second, the storage is also informationally coupled to the receptor, hence the regulatory mechanism is tangled with the external signal and all entries of the MIMMO are non-zero.}
    \label{fig:adaptation}
\end{figure*}

\section*{Information propagation and emergent directionality}
\noindent Let us start again from basic examples and consider the scheme in Figure \ref{fig:information_propagation}a. This represents a three-layer system in which the structure of each layer, for simplicity, reflects the one in Figure \ref{fig:gillespie_infonet}. Its joint probability distribution is: $p_{1,2,3}(t) = p_{1|3}^{\rm st} \, p_{2}^{\rm st} \, p_{3}^{\rm st, eff}(t)$, and thus the only non-zero element of the MIMMO is $I_{13}$ (see Figure \ref{fig:information_propagation}b-c). This information is due to the explicit dependence on $x_3$ in $p_{1|3}^{\rm st}$ induced and modulated by the feedback interaction of strength $f$. This asymmetric effect provides us with an emergent directionality of information propagation, ascribing to the feedback link the role of generating information from slow to fast layers. Instead, the isolated direct link does not contribute to the information content.

We now slightly modify the system, considering the one in Figure \ref{fig:information_propagation}d. Again, the feedback interaction generates information from $x_3$ and $x_1$. At variance with the previous case, the direct link can now propagate the information generated in the first layer on $x_2$, leading to mutual information among all layers (see Figure \ref{fig:information_propagation}e-f). This effect manifests into the joint probability distribution $p_{1,2,3}(t) = p^{\rm st}_{1|3} \, p^{\rm st, eff}_{2|3} \, p^{\rm st}_{3}(t)$. The dependence on $x_3$ appears in all layers, as expected, but $p^{\rm st}_{1|3}$ and $p^{\rm st, eff}_{2|3}$ are conditionally independent. Indeed, the mutual information among them is not directly created by a feedback, but it stems solely from the third layer and is then propagated, i.e., the layers belong to a mPP.

Finally, we move to the four-layer system in Figure \ref{fig:information_propagation}g. When $f \neq 0$, its joint probability distribution is $p_{1,2,3,4}(t) = p^{\rm st}_{1|2} \, p^{\rm st}_{2|4} \, p^{\rm st, eff}_{3|4} \, p^{\rm st}_4$. Here, information is generated from $x_2$ into $x_1$ and from $x_4$ into $x_2$, and this is reflected in the explicit dependencies in the joint probability distribution. Again, the fact that $2$ and $3$ are conditionally independent signals that information is only propagated through direct links. The same happens between $x_1$ and $x_3$. Nevertheless, when $f = 0$, the only non-zero element of the MIMMO is $I_{12}$. Indeed, although $x_2$ creates information in $x_1$, it cannot propagate to $x_3$, since its dynamics is slower than the regulating layer.

Hence, the topological structure of higher-order interactions fully determines the propagation of information. It is clear that our main conclusions only depend on the general structure of the probability in Eq.~\eqref{eqn:joint_probability}, not on the dynamics of the layer nor the form of the higher-order interactions. As a consequence, we can introduce three basic physical principles of information propagation across timescales: (i) feedback interactions generate information from slow to fast layers; (ii) direct interactions alone do not generate information; (iii) information generated through feedback by a slow layer may be propagated through direct interactions to any faster layer. The first two principles highlight the different roles of feedback and direct interactions in affecting the information content of the system. The third one leads to a natural definition of directionality in the information propagation, despite the symmetry of the MIMMO. At the same time, it constrains the accessibility of any control mechanism in stochastic networks in terms of timescales, unveiling that they create mutual information only when acting as regulatory processes, i.e., to faster layers.

\section*{Multiscale signaling networks}
\noindent As a biophysically relevant application, we focus on signaling networks, which have been extensively studied in different biological and living systems \cite{tu_adaptation, tu_tradeoff, wajant2003tumor, dyn_adapt} and constitute an example of controlled stochastic reaction networks \cite{hilfinger2016constraints, yan2019kinetic}. The random walk in each layer describes now a birth-and-death process:
\begin{equation}
    x_\mu \xrightarrow{\; b_\mu(\{x\}_{\nu\neq \mu})/\tau_\mu \;} x_\mu + 1 \qquad x_\mu \xrightarrow{\; d_\mu(\{x\})/\tau_\mu \;} x_\mu - 1
\end{equation}
Since both $b_\mu$ and $d_\mu$, i.e., respectively the birth and death rate, may depend on the state of other layers, this framework can be immediately translated into the scheme in Figure \ref{fig:sketch}. We model inhibitory and excitatory higher-order interactions as increasing the death and birth rate, respectively. The specific forms of $b_\mu(\{x\}_{\nu\neq \mu})$ and $d_\mu(\{x\}$ are given in the Methods.

In Figure \ref{fig:adaptation}a-b we show two representative three-layer signaling architectures that are well-known to provide minimal models for biological adaptation and adaptive responses \cite{ma2009defining, benda2021neuraladaptation}. A receptor $R$ receives a time-varying input $h(t)$ (here a periodically switching signal) and stimulates in turn a fast response described by a readout population $U$. Such response is regulated by a slow storage population $S$, inhibiting either $U$ (Figure \ref{fig:adaptation}a) or $R$ (Figure \ref{fig:adaptation}b) and acting as an effective memory \cite{celani2011molecular,lan2016information,nicoletti2023information}. The layers have $N_U$, $N_R$, and $N_S$ nodes, respectively, describing the maximum number of units each population may have.

We solve the master equation (Eq.~\eqref{eqn:master_equation}) at all times (see Methods). As expected, we find that both architectures display qualitatively similar dynamical behaviors, with the only difference being whether the receptor population decreases its activity (Figure \ref{fig:adaptation}d) or not (Figure \ref{fig:adaptation}c). In particular, the average system response $\ev{U}$ decreases in time upon repeated stimulation, as the storage increases.

However, the information propagation framework allows us to discriminate between the two architecture through their different physical mechanisms that underlie adaptive behavior. In the first case (Figure \ref{fig:adaptation}e), there is no mutual information between receptor and storage, $I_{RS} = 0$ since information is generated solely in the readout $U$. Thus, the regulatory mechanism implemented by $S$ is independent of the receptor activity at the information level. On the contrary, in the second architecture (Figure \ref{fig:adaptation}f), all the entries in the MIMMO are non-zero, since all layers are sequentially connected via feedback interactions. The information generated in $R$ by the storage is then propagated to the readout $U$. Thus, the regulatory function of $S$ is dependent on the receptor activity at the information level, a completely different biochemical mechanism with respect to the previous case.

\begin{figure}
    \centering
    \includegraphics[width = \columnwidth]{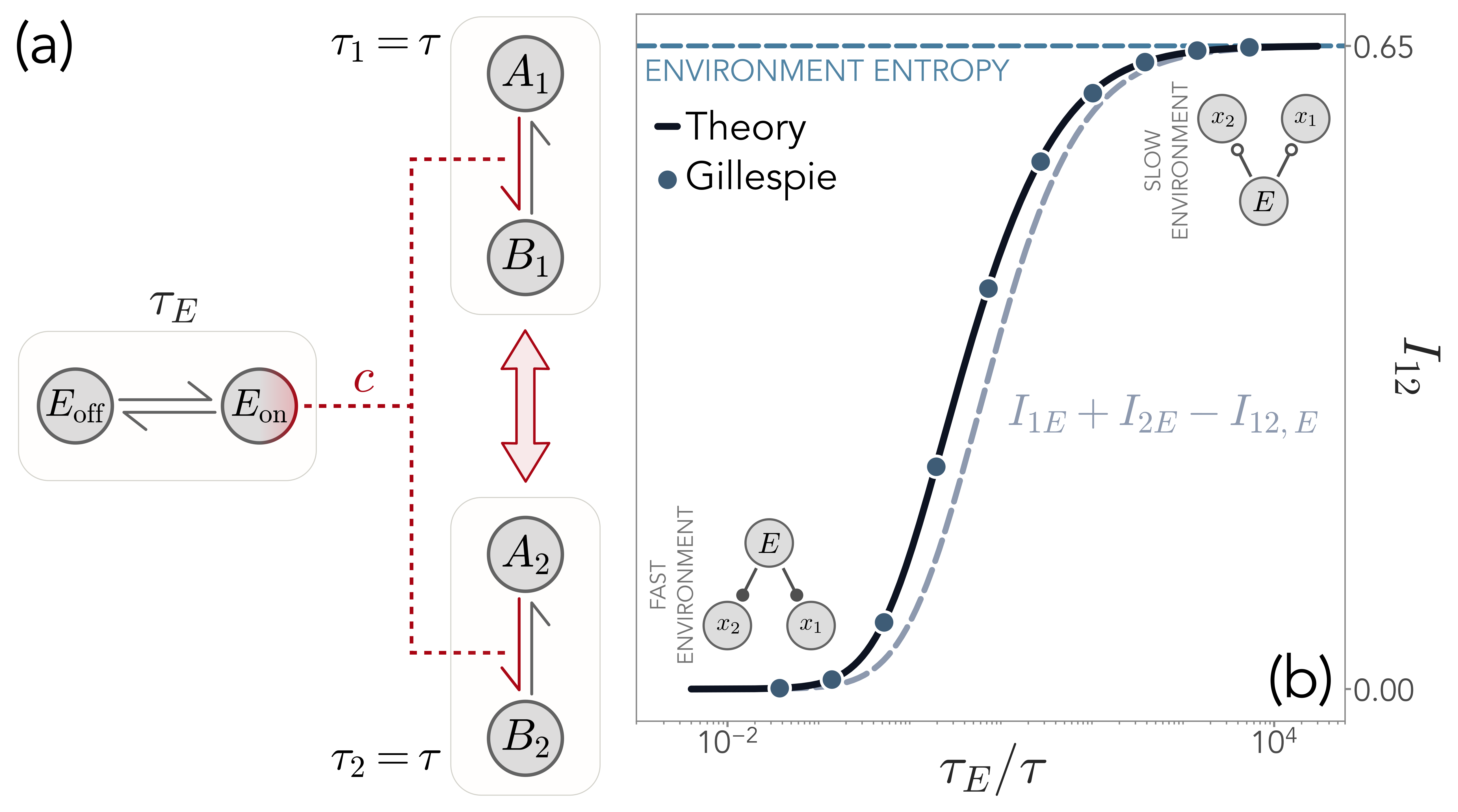}
    \caption{Shared higher-order interactions generate information among random walks in disjoint network components. (a) An archetypal system where one layer has two disjoint components $(A_1, B_2)$ and $(A_2, B_2)$, with two random walks evolving with timescales $\tau_1 = \tau_2 = \tau$. The links of both these components are influenced by another layer, $E$, through a triadic interaction $c$ that models an environmental-like dynamics with timescale $\tau_E$ influencing two separate degrees of freedom $x_1$ and $x_2$. (b) In the limit $\tau_E \gg \tau$, the shared environment induces a non-zero mutual information between the two disjoint random walks, $I_{12}$. The black curve shows $I_{12}$ in the limit $c \to \infty$, while blue dots the corresponding Gillespie simulation. The two random walks are conditionally independent in this limit, so their mutual information can be decomposed exactly (gray dashed line) and converges to the Shannon entropy of the environment $E$ (blue dashed line).}
    \label{fig:environment}
\end{figure}

\section*{Effective information between disjoint networks}
\noindent Finally, we study what happens when information is generated into a layer with disjoint network components. Without loss of generality, we resort to the archetypal system in Figure \ref{fig:environment}. Each disjoint component has two nodes and a random walk dynamics evolving on a timescale $\tau$. Triadic interactions stem from another layer with two nodes, $E_\mathrm{off}$ and $E_{\rm on}$, and a random walk dynamics with timescale $\tau_E$. In particular, $c$ enhances both the transitions $A_1 \to B_1$ and $A_2 \to B_2$ (see Methods). This scenario can be thought of as a model for two distinct degrees of freedom evolving under the influence of a shared environment that exhibits an independent dynamics. Similar settings have been extensively studied in different contexts, from neuroscience to ecology and chemical reaction networks \cite{villa2019bet,berton2020thermodynamics, mariani2022disentangling, nicoletti2021mutual, nicoletti2022mutual,nicoletti2022information,chechkin2017brownian,busiello2021dissipation}.

In the limit $\tau_E \ll \tau$, the shared triadic interaction becomes a direct link and, as such, does not generate mutual information in the system. Conversely, when $\tau_E \ll \tau$, there is a feedback link from $E$ to the two disjoint components at once, so that $I_{1E}$ and $I_{2E}$ are both positive. As a non-trivial consequence, this feedback interaction induces a non-zero mutual information between the two disjoint components, $x_1$ and $x_2$, as well (see Figure \ref{fig:environment}b). This phenomenon of dependencies induced by a switching environment has been first highlighted in \cite{nicoletti2021mutual}, but our framework allows us to immediately pinpoint it. Indeed, in the limit of a slow environment, $p_{1,2}(t) = \sum_E p^\mathrm{st}_{1|E} \, p^\mathrm{st}_{2|E} \, p_E(t)$. Thus, the two random walks are conditionally independent and their effective coupling stems solely from the shared environmental dynamics. In particular, $I_{12}$ converges, in this limit, to the Shannon entropy of $E$, i.e., the maximal information content in the environment, and we can find an exact decomposition as in Figure \ref{fig:environment}b.

\section*{Discussions}
\noindent In this work, we characterized how information propagates and couples the layers of a multilayer network. Although we represented each layer as a physically separated process acting on different timescales and affecting each other through triadic higher-order interactions, our results do not rely on the specific form of inter-layer interactions. Hence, they can be applied to standard multilayer systems with pairwise links across layers (see Methods). In fact, multiple subparts of a large number of biological and non-biological systems are characterized by different timescales, and thus can be seen as layers of a multilayer network \cite{boccaletti2014structure,dedomenico2013mathematical,aleta2019multilayer,bianconi2018multilayer}. Analogously, even if only master equation dynamics have been discussed extensively, the extension to continuous systems is straightforward without affecting our results.

Because of such generality, our work provides a quantitative framework with applications in a large number of fields. Examples can be found in reaction-diffusion systems \cite{busiello2021dissipation,liang2022emergent,brauns2020phase}, ecological population dynamics \cite{villa2019bet}, brain activity \cite{kiebel2008hierarchy,golesorkhi2021brain,rossi2021invariant}, and complex chemical reaction networks \cite{avanzini2023circuit} (see Methods for details). Moreover, experimental developments have enabled substantial progress in understanding complex systems operating at different timescales through the simultaneous recording of their internal variables. In principle, the MIMMO can be estimated directly from recorded timeseries, and it may provide a valuable tool to shed light on the information structure of the system and how regulatory mechanisms are implemented \cite{ma2009defining, rahi2017oscillatory, hilfinger2016constraints, yan2019kinetic}. As a caveat, the precision of this procedure depends on the accuracy of temporal data.

Ultimately, our framework suggests an additional outlook. Ranking the degrees of freedom of a system by the similarity of their timescales allows for the construction of a coarse-grained multi-layer model. With this perspective, our framework might unveil the intrinsic information structure of the underlying dynamics and highlight how information is shared among coarse-grained degrees of freedom. The resulting metric will capture how dynamical processes propagate information across timescales and provide fundamental insights into the physics of regulatory and controlling mechanisms, both internal and environmentally driven.

\begin{acknowledgments}
\noindent We thank Giacomo Barzon for critical discussions and suggestions.
\end{acknowledgments}

\setcounter{equation}{0}
\makeatletter
\renewcommand{\theequation}{S\arabic{equation}}
\renewcommand{\thefigure}{S\arabic{figure}}
\renewcommand{\thesection}{S\Roman{section}} 
\section*{Methods}

\noindent\textbf{Master equation for the multilayer probability.} 
We start from Eq.~\eqref{eqn:master_equation} and seek a solution of the form
\begin{align}
\label{eqn:methods:solution_form_ts}
    p_{1, \dots, N} = & p_{1, \dots, N}^{(1,0)} + \sum_{\mu = 2}^{N-2}\left(\prod_{\nu = 1}^{\mu -1}\epsilon_\nu\right)p_{1, \dots, N}^{(\mu, 0)} + \nonumber \\
    & + \prod_{\mu = 1}^N \epsilon_\mu p_{1, \dots, N}^{(N,1)} + \mathcal{O}(\epsilon^2)
\end{align}
where $\epsilon_\mu = \tau_\mu /\tau_N$ and $\mathcal{O}(\epsilon^2)$ denotes second-order contributions for any $\mu = 1, \dots, N$. Eq.~\eqref{eqn:methods:solution_form_ts} is the standard form of a timescale separation solution, and can be readily obtained by subsequent expansions in $\epsilon_1, \dots, \epsilon_N$ of the joint probability \cite{bo2017multiple}. In the limit $\epsilon_\mu \to 0$, the solution at the leading order obeys
\begin{equation}
\label{eqn:leading_order_me}
    \partial_t p_{1, \dots, N}^{(1,0)}(t) = \left[\hat{\Omega}_N + \sum_{\mu = 1}^{N-1} \frac{\hat{\Omega}_\mu}{\epsilon_\mu} \right]p_{1, \dots, N}^{(1,0)} + \hat{\Omega}_1 p_{1, \dots, N}^{(2,0)}
\end{equation}
where we rescaled time by the slowest timescale. Notice that the solution does not depend on how inter-layers interactions enter in $\hat{\Omega}_\mu$, so our results generalize immediately to the case of pairwise interactions across layers.\\

\noindent\textbf{Minimal propagation paths.} Consider a directed graph $\mathcal{G}$ where each node represents a layer and links between them are given by interactions between layers. We define a propagation path (PP) on $\mathcal{G}$ as a path from $\mu$ to $\nu$ with $\mu > \nu$ and such that it contains at least one direct interaction, i.e., one edge $\alpha \to \beta$ with $\alpha < \beta$. In Figures \ref{fig:information_propagation}a and \ref{fig:information_propagation}d both $2 \to 3 \to 1$ and $3 \to 1 \to 2$ are PPs, repsectively.

We then define the graph $\mathcal{G}^{(\nu)}(\nu^*)$ as the induced subgraph obtained by removing all nodes $\alpha > \nu$ - i.e., all layers slower than $\nu$ - except $\nu^*$. A propagation path from $\mu$ to $\nu$ is a minimal propagation path (mPP) if it is a PP both in $\mathcal{G}$ and in the induced subgraph $\mathcal{G}^{(\nu)}(\mu)$. In Figures \ref{fig:information_propagation}a and \ref{fig:information_propagation}d, $2 \to 3 \to 1$ is not a mPP, but $3 \to 1 \to 2$ is.\\

\noindent\textbf{Exact solution for the multilayer probability.} 
We solve Eq.~\eqref{eqn:leading_order_me} order-by-order, with each order corresponding to a given layer. The first order is immediately solved by
\begin{equation}
    0 = \hat\Omega_1(\{x\}_{\rightsquigarrow 1})p^\mathrm{st}_{1 | \rho(1)}
\end{equation}
where $\rho(1)$ is the set of all slower layers that connect to the first through a feedback link, $\{x\}_{\rightsquigarrow 1}$ is the set of layers directly connected to the first, and we have that $p_{1, \dots, N}^{(1,0)} = p^\mathrm{st}_{1 | \rho(1)}\,p_{2, \dots, N}^{(1,0)}$. Notice that for the first layer, by definition, $\{x\}_{\rightsquigarrow 1} = \rho(1)$. 

The second order, after a summation over $x_1$, reads
\begin{equation}
    0 = \left[\sum_{x_1} \hat\Omega_2(\{x\}_{\rightsquigarrow 2})p^\mathrm{st}_{1 | \rho(1)}\right]p^{\mathrm{st, eff}}_{2 | \rho(2)}
\end{equation}
where the operator in the bracket is the effective operator $\hat\Omega^\mathrm{eff}_{2 | \rho(2)}$. Its stationary probability $p^{\mathrm{st, eff}}_{2 | \rho(2)}$ inherits the dependence on slow layers through both the set $\{\rightsquigarrow x_2\}$ and the dependencies of faster layers - in this case, $\rho(1)$ appearing in $p^\mathrm{st}_{1 | \rho(1)}$. Therefore, the conditional dependencies contained in $\rho(2)$ now include all slower layers that are connected to $2$ through higher-order interactions, as well as all slower layers directly connected to $1$ if there is a direct interaction from $1$ to $2$. Otherwise, in the absence of such directed connection, $\{x\}_{\rightsquigarrow 2}$ does not include $x_1$, and thus the effective operator coincides with $\hat\Omega_2$.

By recursively solving each order and marginalizing over the slower layers, we obtain Eq.~\eqref{eqn:joint_probability} and the effective operators in Eq.~\eqref{eqn:effective_operators}. In particular, the set
\begin{equation}
    \rho(\mu) = \left\{ \nu >\mu : \exists \text{ mPP } \nu \to \mu \text{ or } \hat{C}_{\nu\mu} \ne 0\right\}
\end{equation}
is the set of all layers connected to $\mu$ wither via a mPP or a single feedback link, and it determines the conditional structure of each term of the joint probability $p_{1, \dots, N}(t)$. Let us stress that, by construction, the only time dependence arises at order $\mathcal{O}(1)$, and thus it is ascribed to the effective operator of the slowest layer $\hat\Omega^\mathrm{eff}_{N}$. Notice that, although $\rho(N)$ is an empty set by definition, such an effective operator depends on all the previous layers by which it is influenced through the marginalization in Eq.~\eqref{eqn:effective_operators}.\\

\noindent\textbf{Mutual information between layers.}
Mutual information between any two layers is defined as
\begin{equation}
\label{eqn:methods_mutual}
    I_{\mu \nu} = \sum_{x_\mu \in \mathcal{N}_\mu}\sum_{x_\nu \in \mathcal{N}_\nu} p_{\mu\nu}(x_\mu, x_\nu) \log_2\frac{p_{\mu\nu}(x_\mu, x_\nu)}{p_{\mu}(x_\mu)\, p_\nu(x_\nu)}
\end{equation}
where we explicitly denoted the dependence of the probabilities on the state and the marginal probabilities can be readily computed from the multilayer distribution $p_{1,\dots,N}$. In general, $I_{\mu\nu}$ will depend on time, but in the present work we focused on its stationary limit. Eq.~\eqref{eqn:methods_mutual} can be readily applied to any stochastic process on networks by replacing sums with integrals and the layer state with a state vector $\vec{x}_\mu = (x_\mu^{(1)}, \dots, x_\mu^{(M_\mu)})$, namely
\begin{equation}
    I_{\mu \nu} = \int d\vec{x}_\mu d\vec{x}_\nu \, p_{\mu\nu}(\vec{x}_\mu, \vec{x}_\nu) \log_2\frac{p_{\mu\nu}(\vec{x}_\mu, \vec{x}_\nu)}{p_{\mu}(\vec{x}_\mu)\, p_\nu(\vec{x}_\nu)}
\end{equation}
where the probabilities are probability density functions.\\

\noindent\textbf{Binary triadic interactions.} In Figure \ref{fig:gillespie_infonet} and Figure \ref{fig:information_propagation} we considered, without loss of generality, systems where each layer is a simple two-node network. Denoting such nodes with $\mathcal{N}_\mu = \{A_\mu, B_\mu\}$ for all layers, we take triadic interactions to be
\begin{equation}
    \hat\Omega_\mu = (\hat{W}_\mu)^{i \to j} + \sum_{\nu \ne \mu} \sum_{k \in \mathcal{N}_\nu} C_{\nu\mu}^{k, i \to j} \delta(x_\nu, k)
\end{equation}
so that the transition rates of a layer - i.e., its adjacency matrix - change in a binary fashion depending on the state of another layer, and acting additively with respect to pairwise ones. For simplicity, we set $C_{\mu\nu}^{k, i \to j} = C_{\mu\nu}^\mathrm{eq}\delta_{jk} + C_{\mu\nu}^\mathrm{cr}(1-\delta_{jk})$, where $C_{\mu\nu}^\mathrm{eq}$ represents the triadic interactions of nodes influencing transitions to the same nodes - e.g., $A_\mu$ favoring the transition to $A_\nu$ in another layer - and $C_{\mu\nu}^\mathrm{cr}$ interactions with the opposite effect - e.g., $A_\mu$ favoring the transition to $B_\nu$. We stress that this specific choice does not affect the structure of information, which is fully determined by the conditional structure of Eq.~\eqref{eqn:joint_probability} and thus is independent of the details of the dynamics.\\

\noindent\textbf{Multilayer solution for birth-and-death processes.} A birth-and-death process of a species $X$ may be interpreted as a random walk on a network, with each node representing a given number of particles $x$. In particular, in the case of a finite reservoir with birth rate $b$ and death rate $d$, we consider a network with $N_X$ nodes and adjacency matrix $W_{ij} = \delta_{i,x}[(N_X - x)b \,\delta_{j,x+1} + x d \,\delta_{j,x-1}]$, with appropriate boundary conditions. This formally corresponds to the microscopic reactions
\begin{equation*}
    X \xrightarrow{d} \varnothing_X, \qquad \varnothing_X \xrightarrow{b} X
\end{equation*}
where $\varnothing_X$ represents the finite reservoir. Then, the formalism presented in the main text can be straightforwardly applied. In particular, a physically and biologically meaningful implementation of triadic interactions in this scenario is to take $b_\mu$ and $d_\mu$ of a given layer to depend on the concentration of the number of particles in another layer. That is,
\begin{equation}
\label{eqn:methods:birth_death_rates}
    b_\mu = b^{(0)}_\mu + \sum_{\nu \ne \mu } C_{\nu\mu}^\mathrm{ex}\frac{x_\nu}{N_\nu}, \quad d_\mu = d^{(0)}_\mu + \sum_{\nu \ne \mu } C_{\nu\mu}^\mathrm{in}\frac{x_\nu}{N_\nu}
\end{equation}
where $b^{(0)}_\mu$ and $d^{(0)}_\mu$ are the baseline birth and death rate of layer $\mu$. Thus, in this case, triadic interactions affect all edges of another layer at once, solely distinguishing between birth and death transition rates. In particular, we consider inhibitory interactions $C_{\nu\mu}^\mathrm{in}$ those that increase the death rate, and excitatory interactions $C_{\nu\mu}^\mathrm{in}$ those that increase the birth rate. Since the resulting interactions are linear, the effective operators in Eq.~\eqref{eqn:effective_operators} only depend on the average of the probability distributions $p_{\mu | \rho(\mu)}^{\mathrm{st}(\rightsquigarrow \mu)}$.

Although the solution to a standard birth-and-death process with a finite reservoir and linear rates can be found analytically and is a binomial distribution \cite{gardiner}, the effective operators in Eq.~\eqref{eqn:effective_operators} do not admit a general closed-form solution. Thus, we introduce an efficient numerical scheme to obtain the solution to the master equation at all times. We write the solution in the recursive form
\begin{align}
    p_{1, \dots, N}(t + \Delta t) = & \prod_{\nu = 1}^{N-1}p_{\mu | \rho(\mu)}^{\mathrm{st}(\rightsquigarrow \mu)} \times \\
    & \sum_{\tilde{x}_N} P_N(x_N, t + \Delta t ; \tilde{x}_N, t)p_N^{(\rightsquigarrow N)}(\tilde{x}_N, t) \nonumber
\end{align}
where $\Delta t$ is the timestep and $P_N(x_N, t + \Delta t ; \tilde{x}_N, t)$ is the propagator associated to the effective operator $\hat{\Omega}^\mathrm{eff}_{N}$, which we can write as
\begin{equation}
    P_N(x_N, t + \Delta t ; \tilde{x}_N, t) = p_N^{\mathrm{st}(\rightsquigarrow N)} + \sum_{i = 2}^N \vec\omega_i a^{(i)} e^{\lambda_i \Delta t}
\end{equation}
where $p_N^{\mathrm{st}(\rightsquigarrow N)}$ is the stationary probability of the $N$ layer, and $\vec\omega_i$ and $\lambda_i$ are respectively the eigenvectors and eigenvalues of $\hat{\Omega}^\mathrm{eff}_{N}$, ordered in such a way that $\lambda_0 = 0$. This recursive form is by definition exact when $\Delta t \ll 1$, and it is particularly useful when the input to the system $h(t)$ varies with time, since all effective operators would depend explicitly on time as well.\\

\noindent\textbf{Signaling architectures.} The architecture depicted in Figure \ref{fig:adaptation}a corresponds to the following reactions. The receptor receives a time-varying input $h(t)$, so that
\begin{equation}
    b_{R,1} = b_R^{(0)} + h(t), \quad d_{R,1} = d_R^{(0)}
\end{equation}
with $b_R^{(0)}$ and $d_R^{(0)}$ the baseline birth and death rate. We typically set $b_R^{(0)} = 0$, so the receptor is entirely input-driven. The readout population is stimulated by the receptor and inhibited by the storage, so that
\begin{equation}
    b_{U,1} = b_U^{(0)} + C_{RU}^\mathrm{ex}\frac{r}{N_R}, \quad d_{U,1} = d_U^{(0)} + C_{SU}^\mathrm{in}\frac{s}{N_S}
\end{equation}
with $r$ and $s$ the number of receptor and storage particles, i.e., the node in the corresponding layer. Finally, the storage follows
\begin{equation}
    b_{S,1} = b_S^{(0)} + C_{US}^\mathrm{ex}\frac{u}{N_U}, \quad d_{S,1} = d_S^{(0)}
\end{equation}
so that it is excited by the number of readout units $u$.

The second architecture in Figure \ref{fig:adaptation}b is specified by
\begin{equation}
    b_{R,1} = b_R^{(0)} + h(t), \quad d_{R,1} = d_R^{(0)} + C_{RS}^\mathrm{in}\frac{s}{NS}
\end{equation}
for the receptor, which is inhibited by the storage population. The readout layer is simply given by
\begin{equation}
    b_{U,1} = b_U^{(0)} + C_{RU}^\mathrm{ex}\frac{r}{N_R}\, d_{U,1} = d_U^{(0)}
\end{equation}
and, finally, 
\begin{equation}
    b_{S,1} = b_S^{(0)} + C_{RS}^\mathrm{ex}\frac{r}{N_R}, \quad d_{S,1} = d_S^{(0)}
\end{equation}
so that the storage is excited directly by the receptor.\\

\noindent\textbf{Processes on disjointed components.} We consider a two-layer network, where one layer has two disjointed components with independent states $x_1$ and $x_2$, each with two nodes $(A,B)$. The other layer, with state $E$ and nodes $(E_\mathrm{off}, E_\mathrm{on})$, influences both components at the same time. The rates are $\omega_1^{A \to B} = \omega_2^{A \to B} = (w_B + c\, \delta(E, E_\mathrm{on}))/\tau$, $\omega_1^{B \to A} = \omega_2^{B \to A} = w_A$, $\omega_E^{\mathrm{on} \to \mathrm{off}} = w_0/\tau_E$, $\omega_E^{\mathrm{off} \to \mathrm{on}} = w_e/\tau_E$.

When $\tau_E \gg \tau$, the inter-layer interaction is of the feedback type, so that $p_{1,2,E}(t) = p^\mathrm{st}_{1|E} \, p^\mathrm{st}_{2|E} \, p_E(t)$. In this case, we can write the sum of the mutual information
\begin{align}
    I_{12} + I_{12,E} & = \sum_{1, 2, E} p_{1,2,E} \left[\log\frac{p_{1,2}}{p_1p_2} + \log\frac{p_{1|E}p_{2|E}p_E}{p_{1,2}p_E}\right] \nonumber \\
    & = I_{1,E} + I_{2,E}
\end{align}
which shows that in this limit the mutual information $I_{1,2}$ induced by the shared influence of $E$ is equal to the sum of the information generated by $E$ in $1$ and $2$ minus the information with $E$ that both components have.\\

\noindent\textbf{Triadic interactions at play.} Several biological and non-biological systems exhibit triadic interactions or develop them as a consequence of the presence of different timescales. Here, we provide some examples.

Complex reaction networks under stochastic control are usually modeled by the following reaction scheme:
\begin{equation*}
    x_\mu \xrightarrow{\; r^+_\mu(\{x\})/\tau_\mu \;} x_\mu + b_\mu \qquad x_\mu \xrightarrow{\; r^-_\mu(\{x\})/\tau_\mu \;} x_\mu - 1 \;.
\end{equation*}
This setting is identical to the one implemented to describe signaling networks where non-linear dependencies stem from regulatory species and to the general model of catalytic reaction networks \cite{hilfinger2016constraints,yan2019kinetic}. Further, higher-order interactions in ecological systems, proven to be crucial for stability, share analogous modeling features \cite{grilli2017higher}.

Reaction-diffusion systems in the fast-diffusion regime are described by a master equation \cite{busiello2021dissipation,liang2022emergent}:
\begin{equation*}
    \partial_t P_i = \sum_j W^{\rm eff}_{ij}(x) P_j \;,
\end{equation*}
where $P_i$ is the probability of state $i$ and $\hat{W}^{\rm eff}(x)$ a space-dependent effective operator. Here, the role of the controlling variable is played by the space. These induced dependencies are natural consequences of time-scale separation procedures, hence they arise in identical form in complex chemical networks \cite{avanzini2023circuit}, brain dynamics \cite{mariani2022disentangling}, and stochastic processes with a shared environment \cite{nicoletti2021mutual}.

%

\end{document}